# Multiscale Topology of the Spectroscopic Mixing Space:  Impervious Substrates

Christopher Small[1] and Daniel Sousa[2]

[1] Lamont Doherty Earth Observatory, Columbia University, Palisades, NY 10964. csmall@columbia.edu
[2] Department of Geography, San Diego State University, San Diego, CA, 92182. dan.sousa@sdsu.edu

**Abstract**

Characterization of the topology and dimensionality of spectral feature spaces provides both quantitative and qualitative insight into their information content.  Understanding the characteristics and information content of a spectral feature space is essential to modeling and interpretation of the target properties (landscape, water body, atmospheric column, material surface, etc) of its spectra.  While the larger "global" scale topology of the full feature space may be more influenced by the diversity of continuum shapes and amplitudes of the constituent spectra, the finer "local" scale topology may reflect the diversity of narrowband absorption features present in the spectra.  The reflectance of impervious substrates is of direct relevance to remote sensing of the diversity of built environments worldwide.  Specifically, the ability to distinguish between natural pervious substrates and anthropogenic impervious substrates is a key factor in determining the accuracy with which built environments can be mapped with remotely sensed imagery – particularly given the pervasive spectral mixing that occurs at decameter scales in spectrally heterogeneous built environments.  The objective of this study is to characterize the topology and spectral dimensionality of spectral mixing spaces representing a diversity of built environments in urban areas worldwide using both spaceborne and airborne imaging spectroscopy.  Comparing complementary types of dimensionality reduction to render high dimensional spectral mixing spaces allows for characterization of both spectral dimensionality and mixing space topology.  Both of which are fundamental to choosing appropriate mapping approaches.  Using a diverse collection of 30 decameter-resolution urban core subscenes imaged by the EMIT spaceborne imaging spectrometer and 5 sub-decameter-resolution urban gradient flight lines imaged by the AVIRIS-NG airborne imaging spectrometer, we conduct such characterizations.  Global scale topology of low order principal component-derived mixing spaces resembles the Substrate, Vegetation, Dark (SVD) triangular topology of globally diverse composite mixing spaces found in numerous earlier studies - but these spaces are also characterized by multiple (2-3) distinct substrate endmembers and a wide variety of less common spectra corresponding to synthetic anthropogenic materials.  Local scale topology from 2D and 3D UMAP manifold embeddings reveals the presence of multiple spectrally distinct impervious substrate mixing continua corresponding to geographically distinct built environments.  The ability of EMIT and AVIRIS-NG surface reflectance products to distinguish a variety of spectral continuum shapes and narrowband absorption features in a diverse collection of impervious substrates and spectral mixing continua suggests that they, and similar imaging spectrometers, could be useful tools for mapping spatial extent and changes in built environments worldwide.

**Introduction**

The statistical and topological properties of spectral feature spaces are direct expressions of the populations of spectra they represent.  Specifically, the diversity of spectral continuum shapes and amplitudes and the diversity of absorption features superimposed on them *(Clark and Roush*

*1984), (Hapke 2012)*. Characterization of the topology and dimensionality of spectral feature spaces provides both quantitative and qualitative insight into their information content *(Boardman 1993, 1994)*. Here we follow the primary definition of topology as given by the Oxford English Dictionary; *The way in which constituent parts are interrelated or arranged*. The topology of an object is inherently dependent on its dimensionality. In the case of imaging spectroscopy, the dimensionality of a spectral feature space is dependent on both the intrinsic dimensionality of the population of spectra and on the sampling (both spatial and spectral) of the imaging sensor, among other factors.

Understanding the characteristics and information content of a spectral feature space is essential to modeling and interpretation of the target properties (landscape, water body, atmospheric column, material surface, etc). In the context of this analysis, modeling can include both physically-based continuous models (e.g. spectral mixture models) and categorically-defined discrete classifications (e.g. thematic land cover maps). In both cases, the model represents a lower dimensional realization of a higher dimensional feature space. Characterization of the topology and dimensionality of a spectral feature space can therefore inform the design of appropriate models to represent its information content.

Accurate representation of a spectral feature space often depends on the statistical variance scale of features contained within. While the larger "global" scale topology of the full feature space may be more influenced by the diversity of continuum shapes and amplitudes of the constituent spectra, the finer "local" scale topology may reflect the diversity of narrowband absorption features present in the spectra. Depending on the application of the model, either or both scales may be relevant to how the feature space is modeled. Whereas clustering within the feature space can determine how accurately it can be categorized with a discrete classification, the number of distinct spectral endmembers and linearity of the topology can guide the design of spectral mixture models. The primary focus of this analysis is on the topology and dimensionality of spectral feature spaces of reflectance spectra of impervious substrates. Specifically, those found in spatially heterogeneous built environments.

The reflectance of impervious substrates is of immediate relevance to remote sensing of the diversity of built environments worldwide (e.g. *(Okujeni, Linden, and Hostert 2015), (Weng 2012), (Wang and Li 2015)*). Specifically, the ability to distinguish between natural pervious substrates and anthropogenic impervious substrates is a key factor in determining the accuracy with which built environments can be mapped. Particularly the peripheries where pervious and impervious substrates are comingled across multiple spatial scales. The inability to accurately distinguish between pervious and impervious substrates with broadband optical imagery has confounded efforts to map human settlements for decades *(Small 2002; Small et al. 2018)*. The fundamental question we seek to address is whether narrowband imaging spectroscopy can resolve diagnostic spectral features of impervious substrates in spectral mixtures consistently enough to allow for more accurate mapping of composition and spatial extent.

The overall objective of this analysis is to characterize the topology and spectral dimensionality of spectral feature spaces composed of a diversity of built environments from urban areas worldwide. To achieve this, we construct a composite spectral feature space as a mosaic of 30 urban cores from 24 cities imaged by NASA's EMIT spaceborne imaging spectrometer.

Because EMIT's 60 m Instantaneous Field of View (IFoV) is considerably larger than the 10-20 m characteristic scale of most built environments (Small 2003, 2009), we also compare spectral feature spaces derived from subdecameter (3-8 m) resolution spectra collected by NASA's airborne AVIRIS-NG from three diverse urban cores. While the subdecameter resolution imagery generally oversamples individual features (e.g. buildings, streets, sidewalks), spectrally mixed pixels are nonetheless common due to the multiscale heterogeneity of urban mosaics. Because many subdecameter IFoVs, and most 60 m IFoVs, image spectral mixtures in built environments, we refer to these spectral feature spaces as mixing spaces to acknowledge the ubiquitous impact of spectral mixing to the topology of the feature spaces.

The specific objectives of the analysis are to characterize the differences and similarities of multiple scales of mixing space topology using complementary approaches to dimensionality reduction. With application to continuous spectral mixture models, this characterization will allow for identification of spectral endmembers and dimensionality of viable models. With application to discrete thematic classifications, this characterization will allow for identification of distinct clusters within the mixing space and quantify the spectral separability of such clusters.

**Data**

*EMIT*

The primary source of data for this analysis is a 30-scene mosaic of spectroscopic imagery from NASA's Earth Mineral Dust Source Investigation (EMIT) mission *(Bradley et al. 2020)*. The EMIT instrument is a Dyson imaging spectrometer with an 11° cross-track field of view. EMIT is an optically fast (F/1.8) and wide-swath (1240 samples) optical system achieving roughly 7.4 nm spectral sampling across the 380–2500 nm spectral range at high signal-to-noise (SNR) *(Bradley et al. 2020)* The ground sampling distance of EMIT pixel spectra is 60 m. EMIT was launched via SpaceX Dragon on 14 July 2022, and successfully autonomously docked to the forward-facing port of the International Space Station (ISS) *(LP_DAAC 2023)*. All EMIT data were downloaded from https://search.earthdata.nasa.gov/ as the standard Level-2A ISOFIT-corrected surface reflectance product (EMITL2ARFL v001). Cloud and data quality masks were visually inspected, but not used in this analysis. Default bad bands lists (bands 128-142 & 188-213) provided with the data were applied. All EMIT data and algorithms are fully open source.

*AVIRIS-NG*

Archival sub-decameter (3 – 8 m) airborne imaging spectroscopy data were used to supplement EMIT data for Kolkata, London, and Los Angeles. Airborne data analysis was based on 5 low altitude flight lines collected between 2018 and 2021 by the Advanced Visible Infrared Imaging Spectrometer-Next Generation (AVIRIS-NG) instrument. AVIRIS-NG measures radiance from 380 to 2510 nm at 5 nm intervals *(Chapman et al. 2019)*. All flight lines were downloaded from the AVIRIS-NG data portal (https://avirisng.jpl.nasa.gov/dataportal, accessed on 1 June 2023) as orthorectified radiance. Each line was converted to surface reflectance using the Imaging Spectrometer Optimal FITting algorithm (ISOFIT) *(Thompson et al. 2018)*. ISOFIT version 2.9.2 was used, as cloned from https://github.com/isofit/isofit (accessed on 1 June 2022). For all ISOFIT runs, the empirical line method flag was turned on (ELM = 1), with segmentation size = 200.

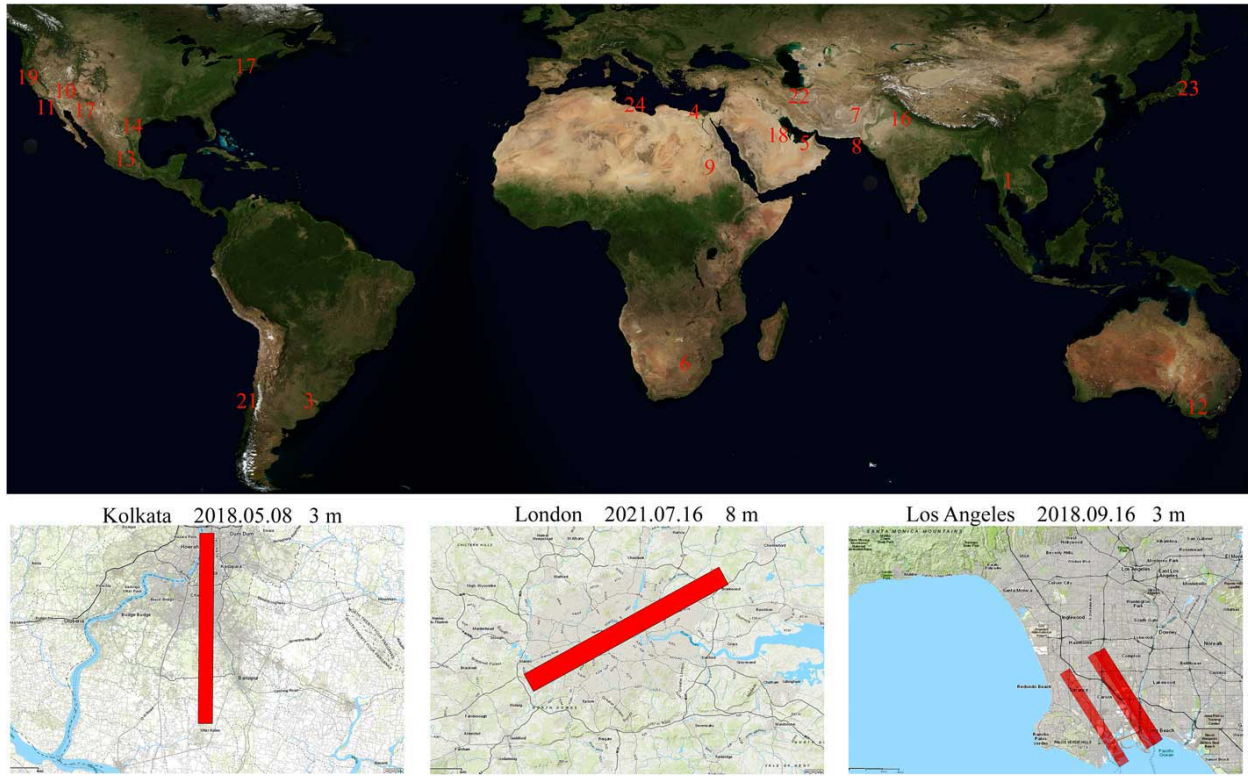

*Figure 1 Index maps for urban EMIT scenes (top) and AVIRIS-NG lines (bottom). The AVIRIS-NG swath is ~2 km wide for Kolkata (1 line) and Los Angeles (3 lines) and ~5.5 km for London. All EMIT scenes acquired between 2022.08.17 and 2023.03.27. Index numbers correspond to scene IDs given in Table 1.*

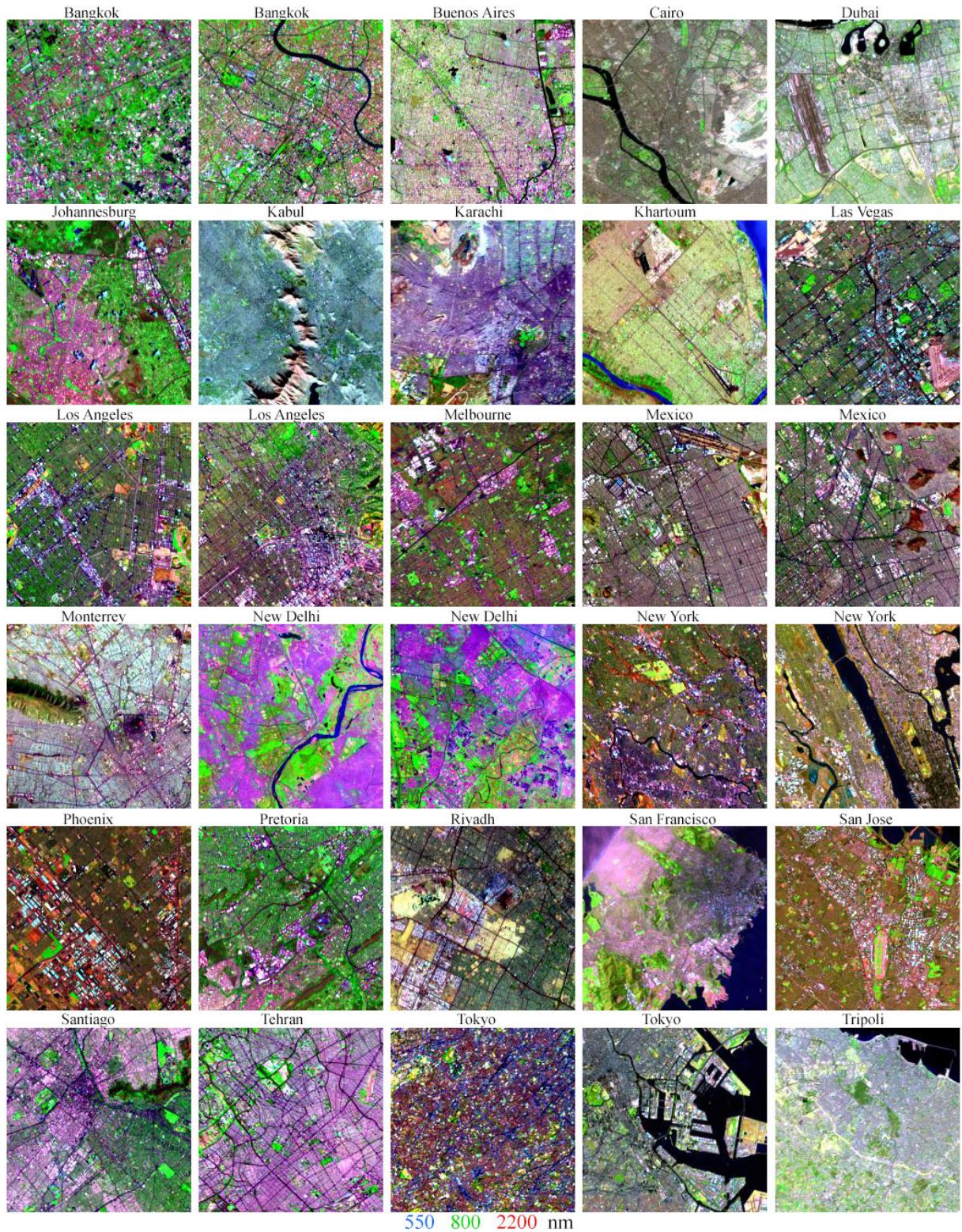

*Figure 2a   EMIT subscenes from 24 urban cores.  Each subscene is 200 x 200 pixel spectra at 60 m resolution (12 x 12 km).  Locations are chosen to maximize the spectral diversity of impervious surfaces.  Subscene-specific 2% linear stretches emphasize intra-scene diversity.*

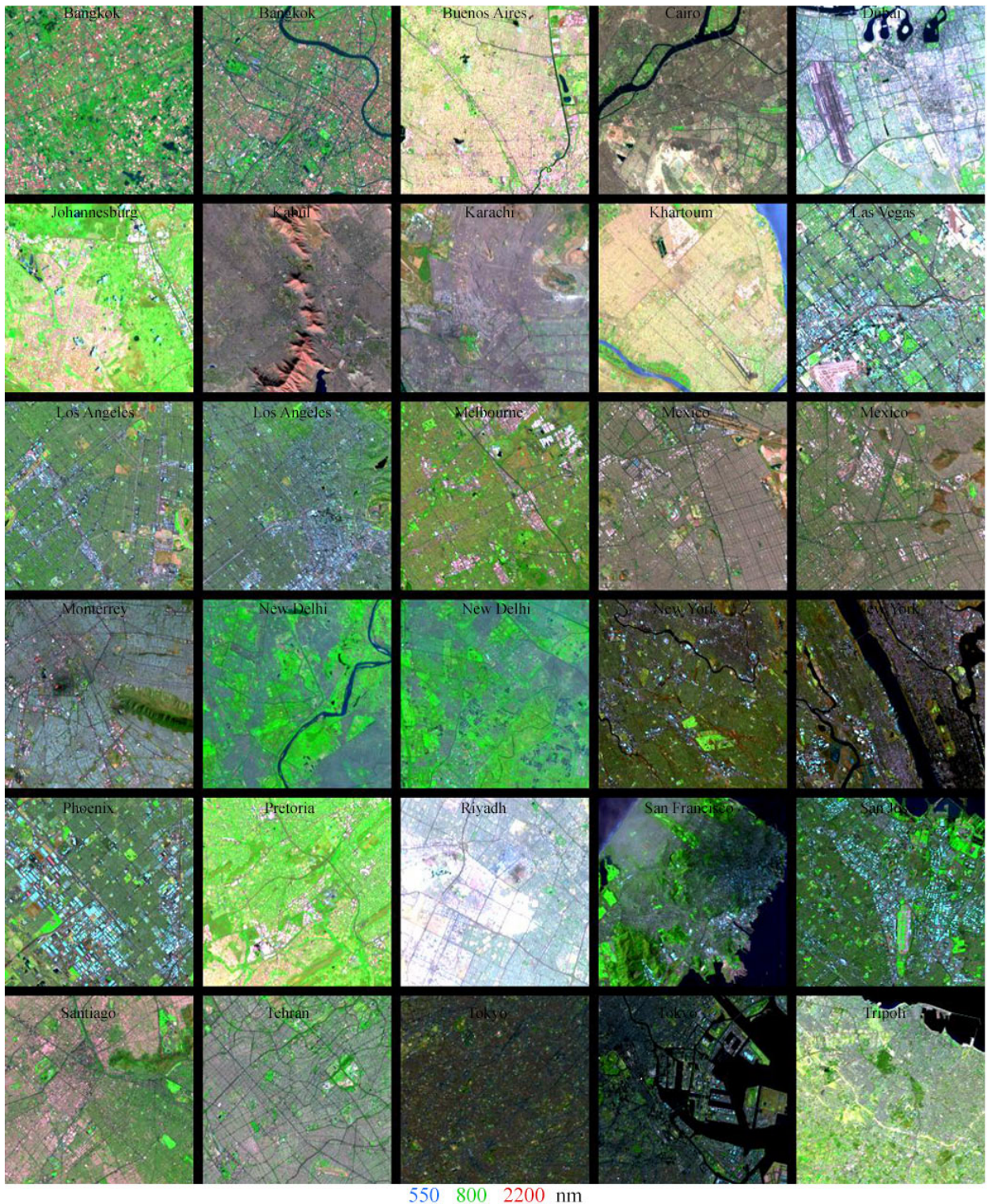

*Figure 2b  EMIT subscenes from 24 urban cores.  A common 2% linear stretch to all scenes emphasizes inter-scene diversity among different cities.*

**Methods**

*Dimensionality Reduction*
In this analysis we use two complementary forms of dimensionality reduction to render low dimensional projections of spectral mixing spaces. Specifically, matrix factorization and neighbor graphs. The principal component (PC) transformation *(Pearson 1901)* is a linear matrix factorization that retrieves the low rank structure of the mixing space by maximizing variance in the minimum number of orthogonal dimensions in which variance diminishes monotonically with increasing dimension. As such, projections of the low dimensional principal components (PCs) render the topology of the mixing space in a way that emphasizes the global structure resulting from the higher variance continuum of each spectrum. In contrast, Uniform Manifold Approximation and Projection (UMAP) *(McInnes et al. 2018)* constructs an adjacency preserving neighbor graph of the high dimensional mixing space and embeds the graph in a low dimensional embedding space in which statistically local structure related to low variance features, like narrowband absorptions, can be resolved. The topology of the low dimensional PC projections is physically interpretable, in part, because these projections reveal the spectral endmembers that are most distinct on the basis of spectral continuum shape bounding the feature space of spectral mixtures of these endmember spectra. The low dimensional embeddings given by UMAP may also preserve some physically interpretable global structure of the mixing space but also reveal finer scale clustering and mixing relations within the mixing space related to low variance features not clearly preserved in the low order PC projections.

For each of the composite mixing spaces described above we compute both a traditional (L2 norm minimization) PC transform, as well as a Robust PCA (RPCA) factorization *(Candès et al. 2011)*. RPCA separates the low rank component of the mixing space from a sparse component containing outliers and transient features that may bias the projections provided by the traditional L2 transformation. A more detailed explanation of RPCA is given in Appendix A. Subsequent Singular Value Decomposition of the low rank (L) and sparse (S) components provides the variance distributions of the orthogonal dimensions of each. For the EMIT mosaic, these variance distributions, described in more detail in Appendix A, reveal that the vast majority of variance is preserved in the low rank component, with the sparse component containing narrowband anomalies adjacent to water absorption features, presumably related to the performance of the atmospheric correction on anomalous spectra. Both the L2 and low rank mixing spaces of the EMIT mosaic are effectively 3D, with > 99% of variance in the three low order dimensions. The topology of the L2 and low rank components are virtually identical, revealing the same spectral endmembers and linearity of mixing.

Both 2D and 3D UMAP embeddings are computed for each of the composite mixing spaces described above. Because the topology of the UMAP embedding can be significantly impacted by both the "nearest neighbor" and "minimum distance" hyperparameters chosen, we conduct a hyperparameter sweep on both to determine the sensitivity of the topology and the consistency of internal structure of the mixing space (Appendix B).

*Topology, Endmembers, Mixture Modeling*
The topology of the low dimensional PC space is a function of the presence and identity of the spectral endmembers bounding the mixing space. In addition, the curvature (i.e., linearity, concavity or convexity) of the convex hull spanning the endmember apexes reveals the linearity (or lack thereof) of spectral mixing among bounding mixtures of endmembers within the space. As such, the dimensionality (quantified by the variance partition of the eigenvalues of the PC factorization) and the topology of the low dimensional space provide a basis for the design of data-adapted linear spectral mixture models. Inversion of the linear spectral mixture model yields endmember fraction estimates for each pixel spectrum within the mixing space. Together, the spatial maps of spectral endmember fractions provide a low dimensional physically-based projection of the higher dimensional spectroscopic mixing space.

*Spectral Separability*
While the PC-derived mixing spaces tend to reflect the global structure of the continuum shapes of the constituent spectra, the local scale topology preserved in the UMAP-derived mixing spaces often shows much greater degrees of internal clustering related to low variance features like narrowband absorptions shared by similar spectra. This clustering may take the form of discrete clusters or continuous tendrils within the mixing space. The presence, or absence, of discrete clusters has immediate implications for discrete thematic classifications which require high degrees of spectral separability among classes. Labeling discrete clusters within UMAP embeddings allows for estimation of spectral separability with metrics such as Transformed Divergence *(Swain 1973)* and Jeffries-Matusita distance *(Richards 1999)*. Even when fully discrete clusters cannot be identified, distinct continua within the mixing space can be interpreted by comparison of spectra corresponding to different continua, or by back-projecting 3D UMAP coordinates into geographic space as RGB composites to reveal spatial continuity and geographic distinction of coherent spatial features with similar color renderings. We use both approaches with the EMIT mosaic to identify compositional consistencies in impervious substrates in different urban settings.

**Results**

Consistent with earlier characterizations of multispectral *(Small 2004; Small and Sousa 2022; Sousa and Small 2017a)* and spectroscopic *(Sousa and Small 2023, 2017b)* mixing spaces, global variance of the 30 scene EMIT mosaic is effectively 3D, with > 99% of spectral variance contained in three low order PC dimensions. Also consistent with previous studies, this mixing space is bounded by Substrate, Vegetation and Dark endmembers with strongly linear mixing along the binary continua radiating from the Dark endmember (Fig. 3). Both Vegetation and Dark endmembers have well-defined apexes, while the three Substrate endmembers are more diffuse, each with a relatively well-defined Interior (I) apex and a constellation of higher amplitude Exterior (E) endmembers around the periphery of the apex (Fig. 3, 4). However, unlike previous studies, the mixing space of the EMIT mosaic also contains two additional endmembers with diffuse interior apexes and constellations of higher amplitude external endmembers (Fig. 3, 4) corresponding to SWIR-absorptive spectra and SWIR-reflective spectra.

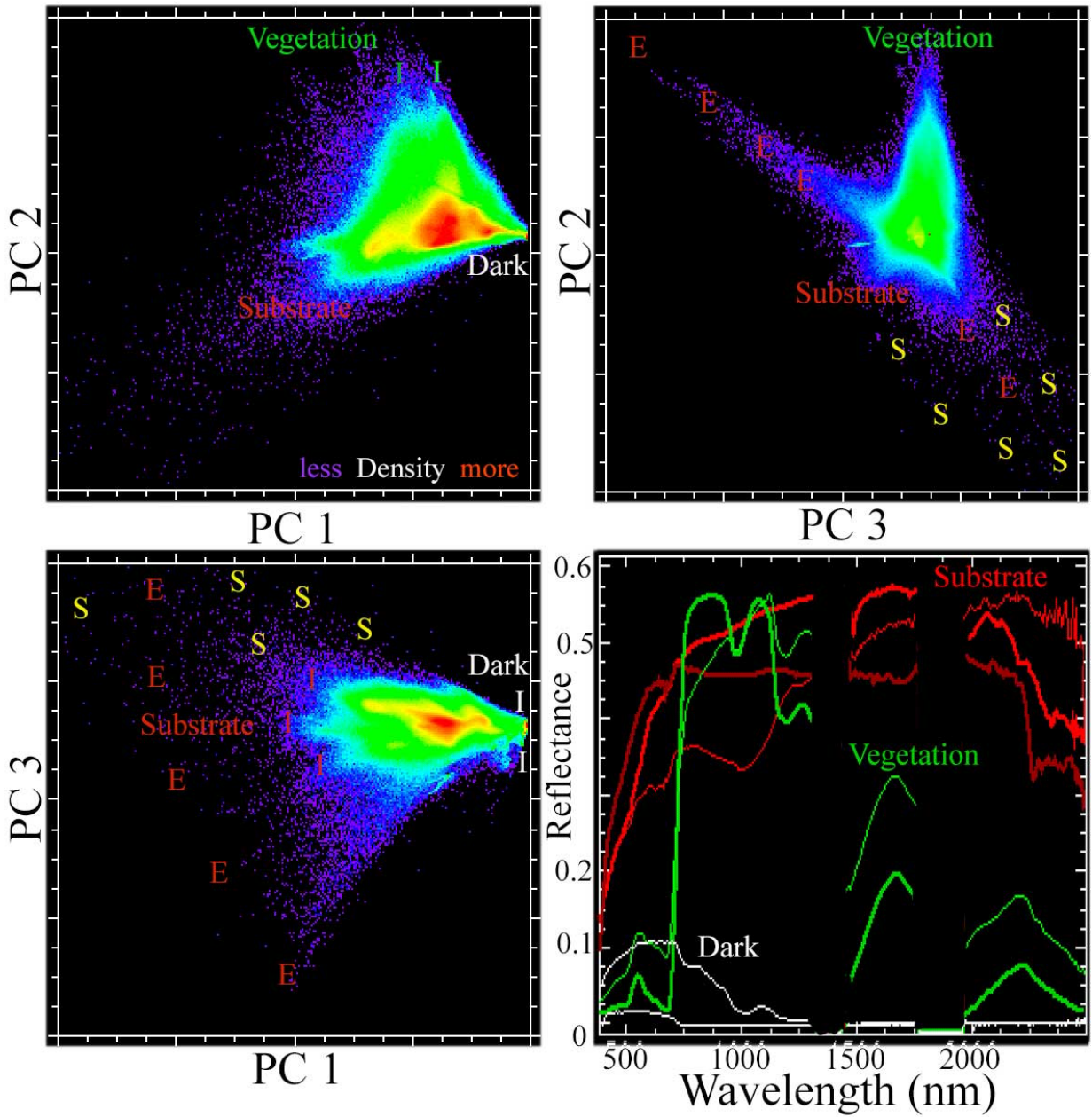

*Figure 3   Spectral mixing space and endmembers for the urban EMIT mosaic.  Three low order PCs represent 98.9% of total variance.  Apexes of the dense cloud correspond to Substrate, Vegetation and Dark (SVD) interior (I) spectral endmembers.  Outlying spectra in the sparse cloud are distinct.  Exterior (E) and specular (S) spectra shown in Figure 4.*

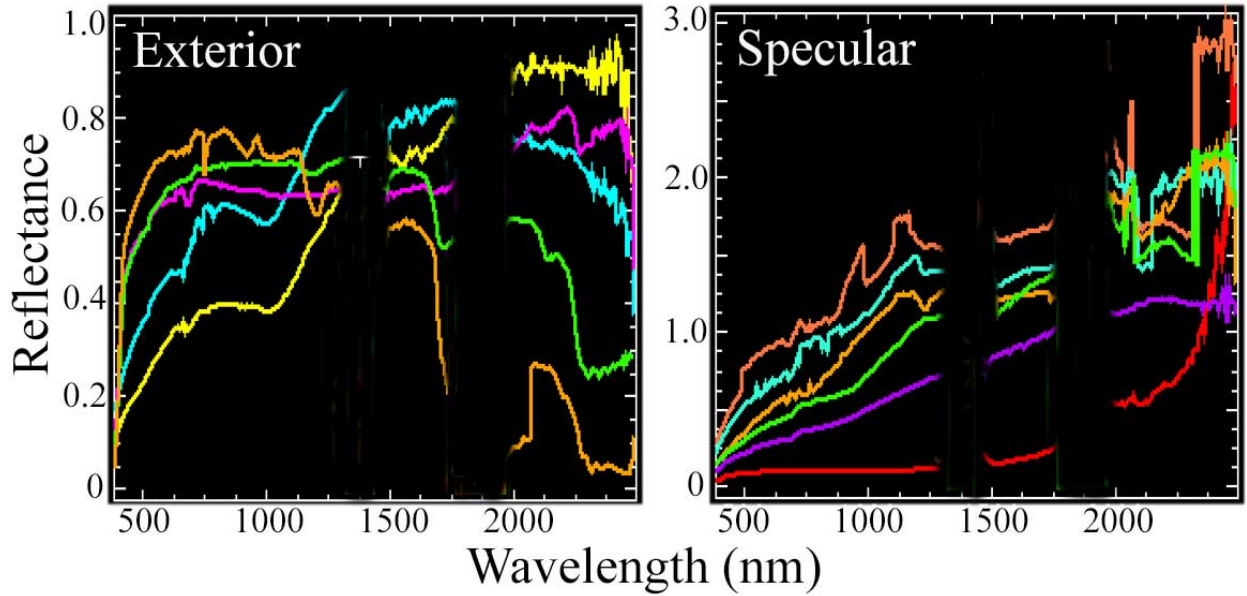

*Figure 4 Exterior Substrate and specular endmembers from the PC mixing space in Fig. 3. Exterior endmembers span a continuum from VNIR-bright/SWIR-dark to VNIR-dark/SWIR-bright. Specular endmembers tend to be SWIR-bright. The red specular endmember is an emission spectrum from a steel mill in Monterrey.*

In contrast, for nearest neighbor distances greater than 5, the UMAP embeddings converge to a consistent topology that preserves the mixing continua spanning a single Dark and multiple Substrate endmembers, while also containing multiple distinct clusters and continua spanning Vegetation and Substrate endmembers (Fig. 5). Within the Dark-Substrate continua are multiple distinct tendrils and mixing continua, each corresponding to a geographically distinct subscene. Spectra sampled along these distinct continua show consistent spectral continuum shape and absorption features, with amplitude varying along Dark-Substrate binary mixing lines (Fig. 6). Similar mixing trends are observed within Substrate-Vegetation clusters and continua. It is noteworthy that two of the distinct Dark-Substrate continua contain multiple geographically proximal cities. Pixel spectra within Dubai+Riyadh and Kabul+Tehran continua are each extensively comingled, in a manner similar to the broad Dark-Substrate continuum containing the remaining subscenes that do not belong to the geographically distinct continua discussed above. Reflectance spectra from all pairs of distinct clusters and continuum ends within the 3D UMAP space in Fig. 5 have Transformed Divergence scores > 1.97, with the vast majority of cluster pairs having scores of 2.0.

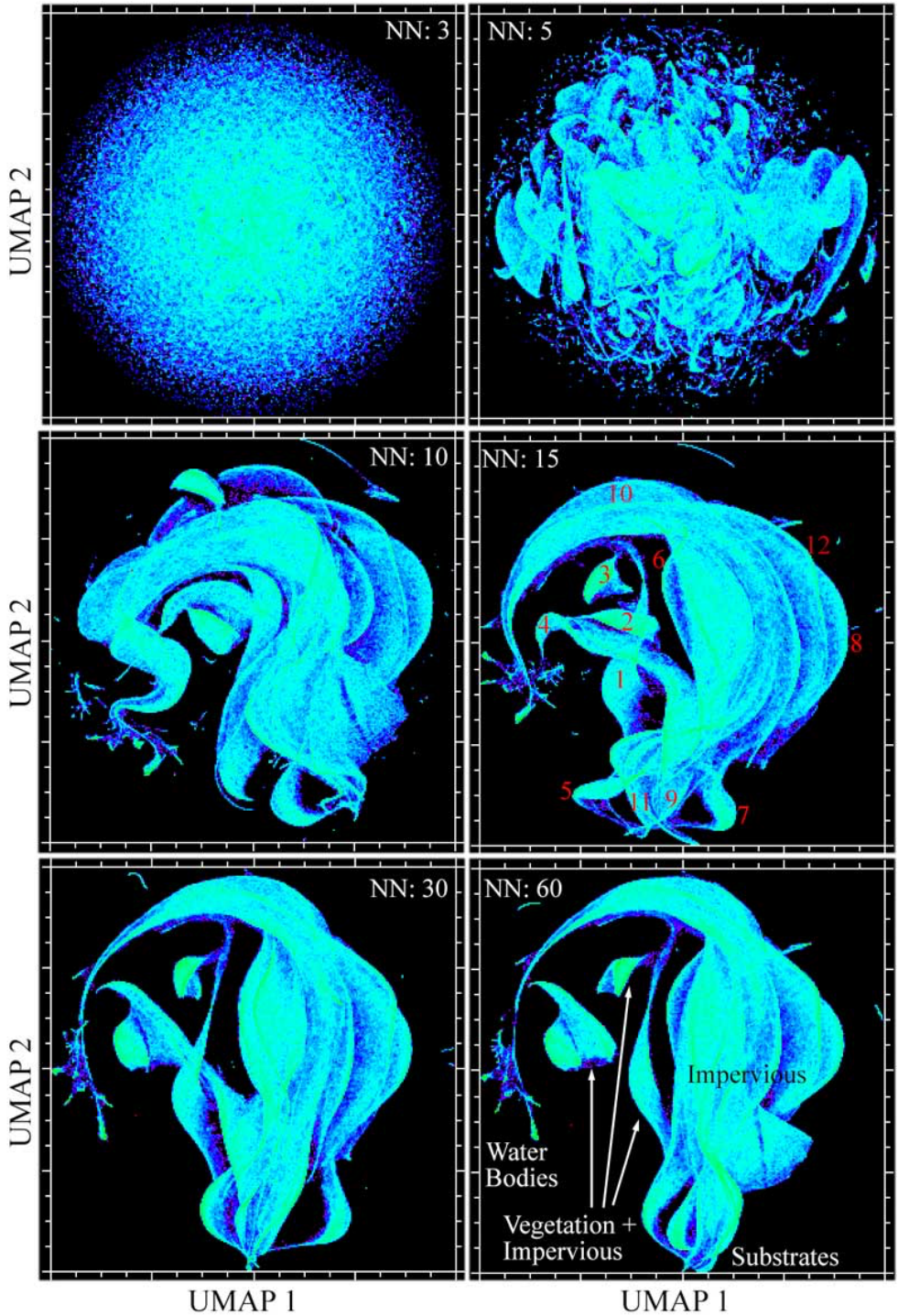

*Figure 5 UMAP embeddings for the urban EMIT spectral mixing space. For NN = 5 a multitude of small clusters reveal spatial variability of water body reflectance amid a larger continuum of impervious substrate and vegetation reflectance. For larger NN settings the topology of the feature space converges to a multi-limbed impervious continuum and a few distinct vegetation+impervious mixture clusters. Index numbers (center right) correspond to reflectance continua of specific cities in Figure 6.*

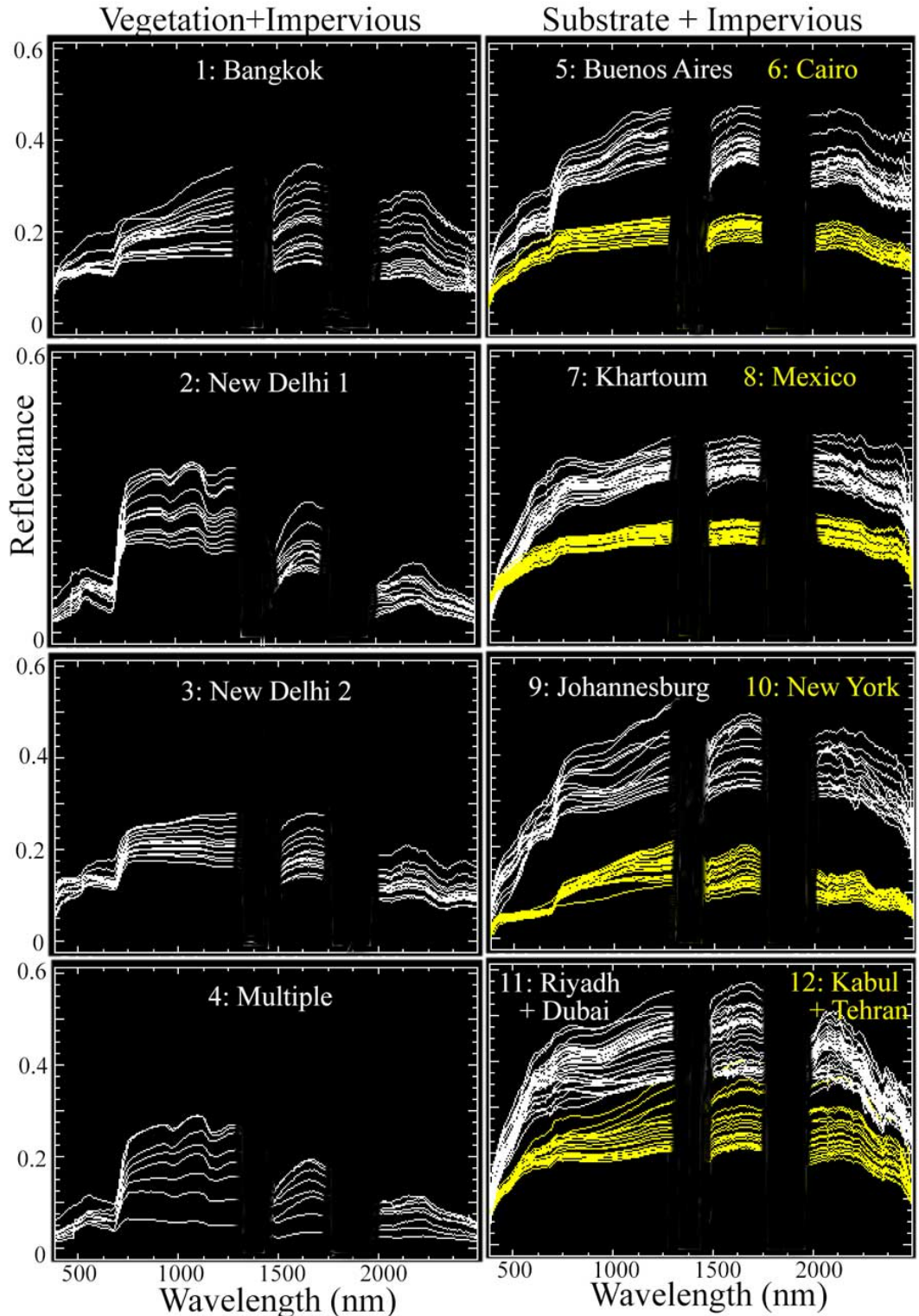

*Figure 6 Reflectance continua corresponding to mixing space continua labeled in Figure 5. The Vegetation+Impervious continuum spanning multiple cities (4) intersects a Substrate + Impervious continuum also spanning multiple cities. In two cases, geographically proximal cities (Riyadh+Dubai and Kabul+Tehran) comingle within single continua.*

The sub-decameter AVIRIS-NG mosaics for Kolkata, London and Los Angeles (Fig. 7) all show the familiar SVD topology, in both the PC and UMAP mixing spaces (Fig. 8). All three show two distinct Substrate endmembers with both interior apex and diffuse constellations. The London mixing space also resolves two distinct vegetation endmembers, corresponding to tree canopy and grass. However, the more compositionally diverse mixing space of the Los Angeles mosaic contains two additional endmembers corresponding to synthetic substrates like paints and plastics. The UMAP embeddings show similar SVD-bounded global topologies for all three mosaics (Fig. 8), but the Los Angeles embedding also contains multiple discrete clusters corresponding to different water bodies and materials whose spectra are distinct from the larger SVD mixing continuum (Fig. 9).

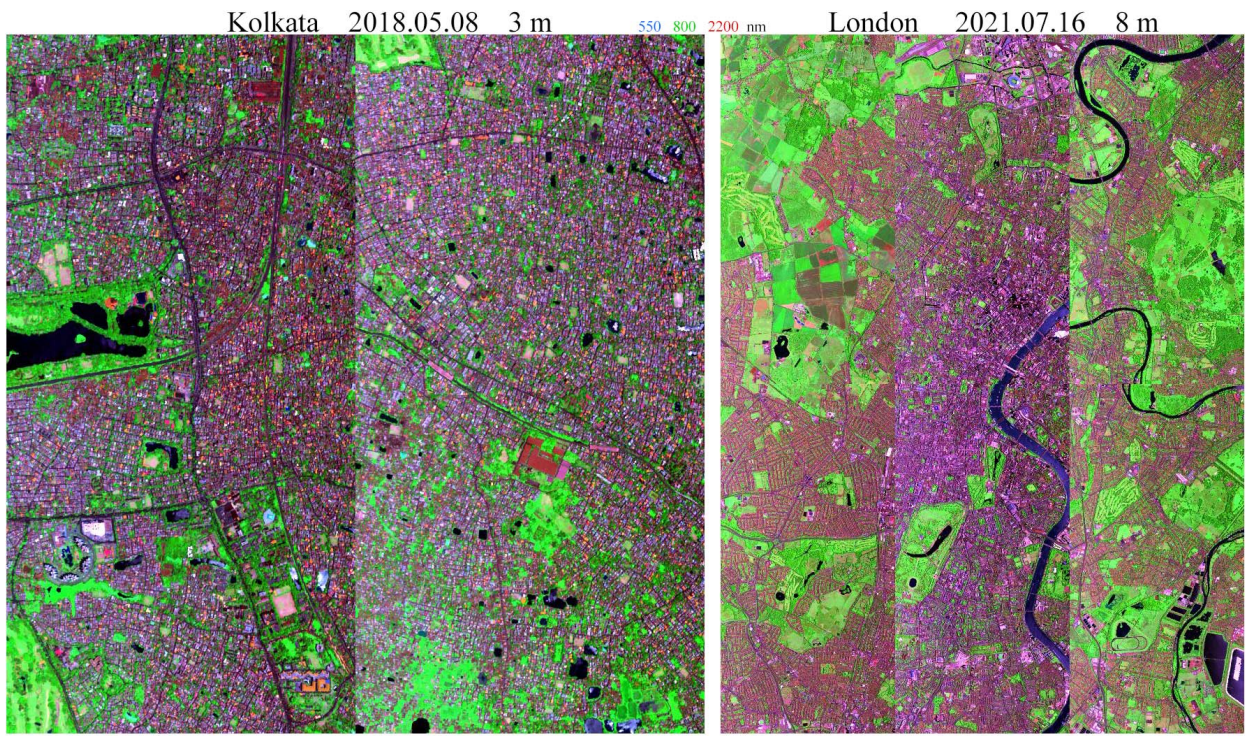

*Figure 7a VNS composites for AVIRIS-NG mosaics of Kolkata and London. Each strip is 500 pixels wide. Despite the higher spatial resolution of the Kolkata line, both resolutions approach the characteristic scale of smaller structures in each city.*

Los Angeles     2018.09.16     3 m

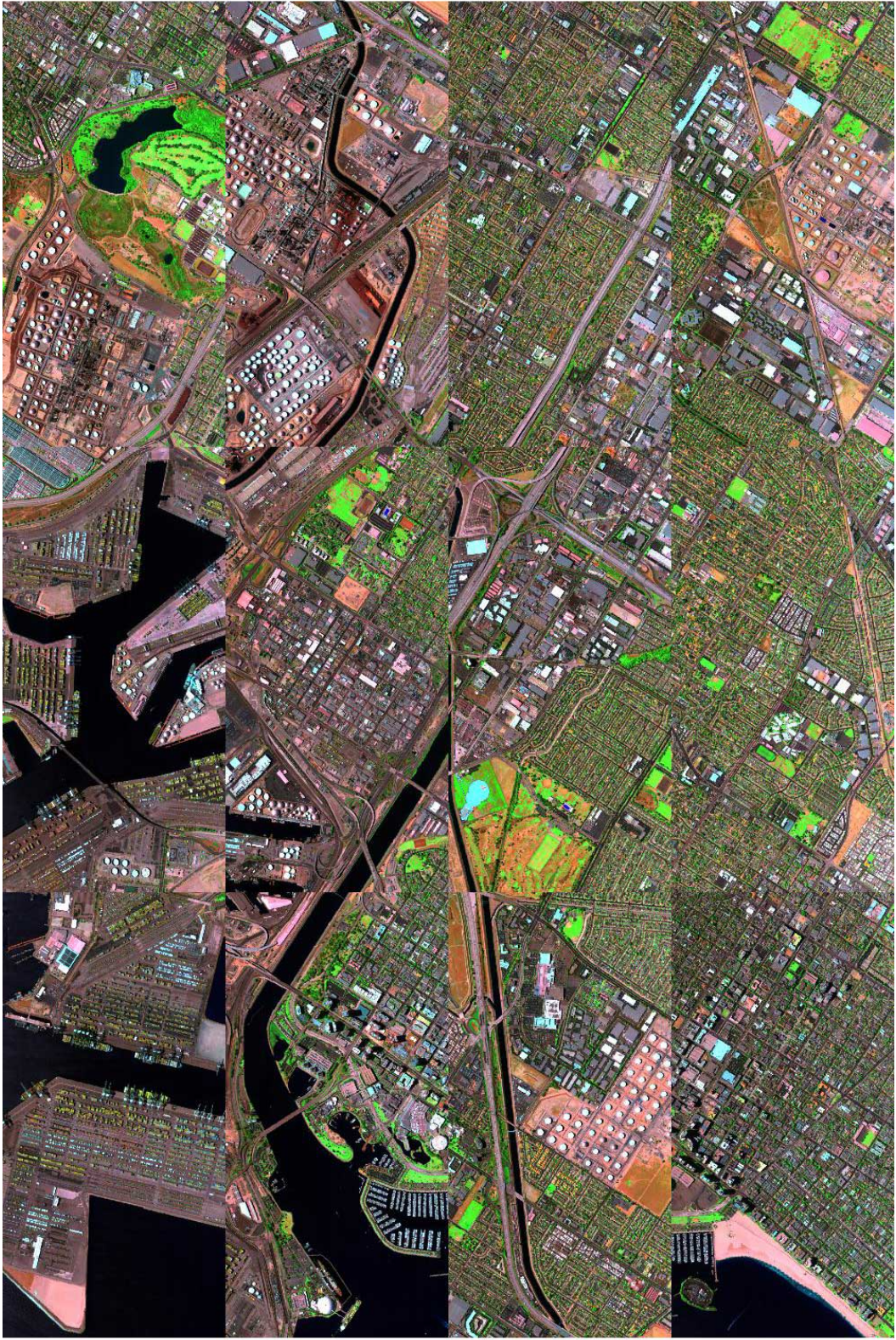

*Figure 7b  VNS composite for AVIRIS-NG mosaic of Los Angeles.  Each strip is 500 pixels wide.*

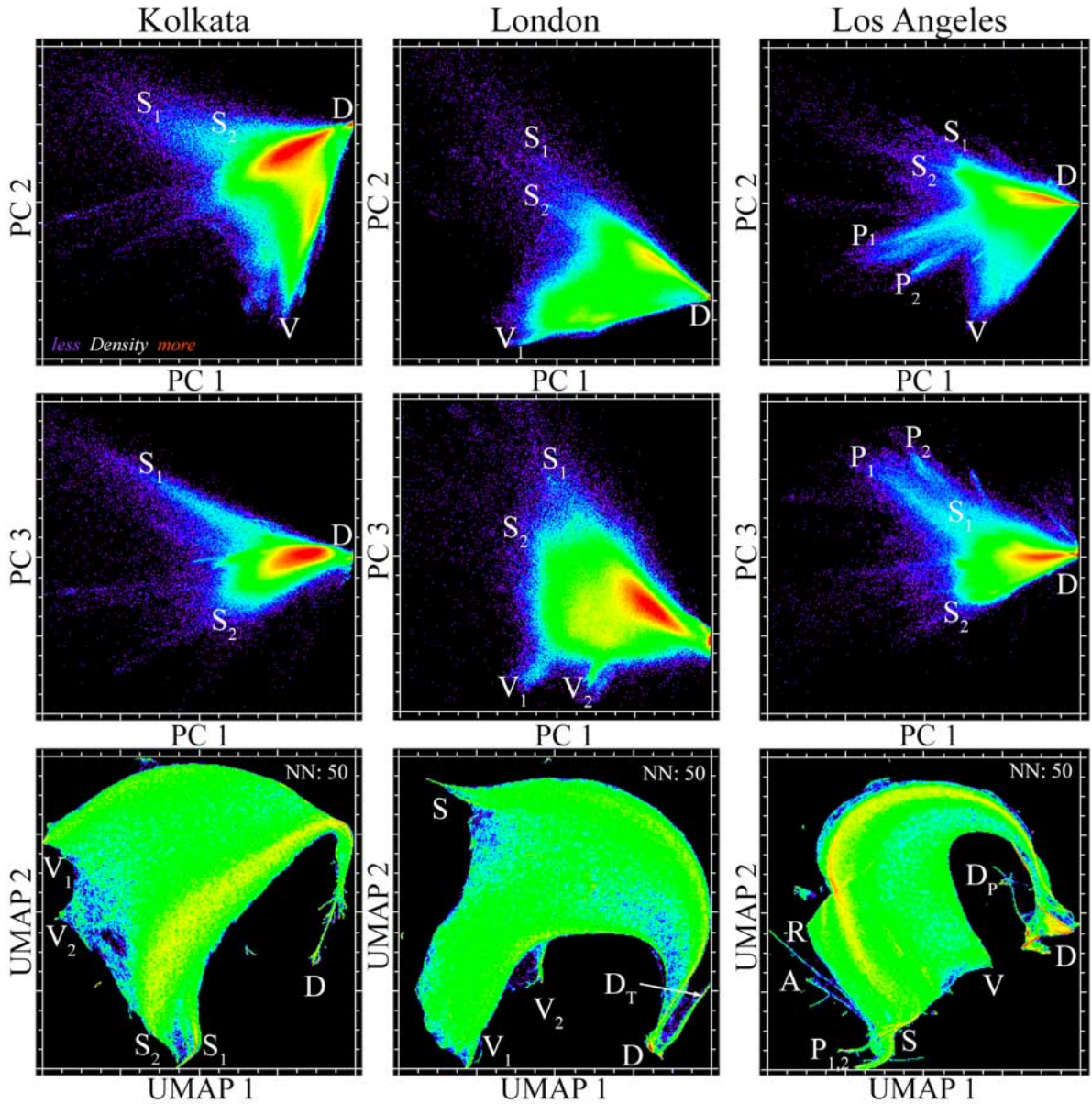

*Figure 8 Spectral mixing spaces for the urban AVIRIS-NG mosaics. Orthogonal projections of the PC spaces (top, center) all show the triangular SVD structure with constellations of exterior substrate EMs. In both PC and UMAP spaces, most impervious surfaces cluster along Substrate-Dark mixing trends (red, yellow). In each space, there are at least 2 interior substrate EMs, with additional roofing (R) and paints (P) in Los Angeles. The SVD structure is also preserved in the UMAP spaces, with anomalous reflectance tendrils such as solar panel arrays (A) and other synthetic material spectra shown in Figure 9. Dark endmember clusters show different water bodies, including Thames River (DT) and pond water (DP).*

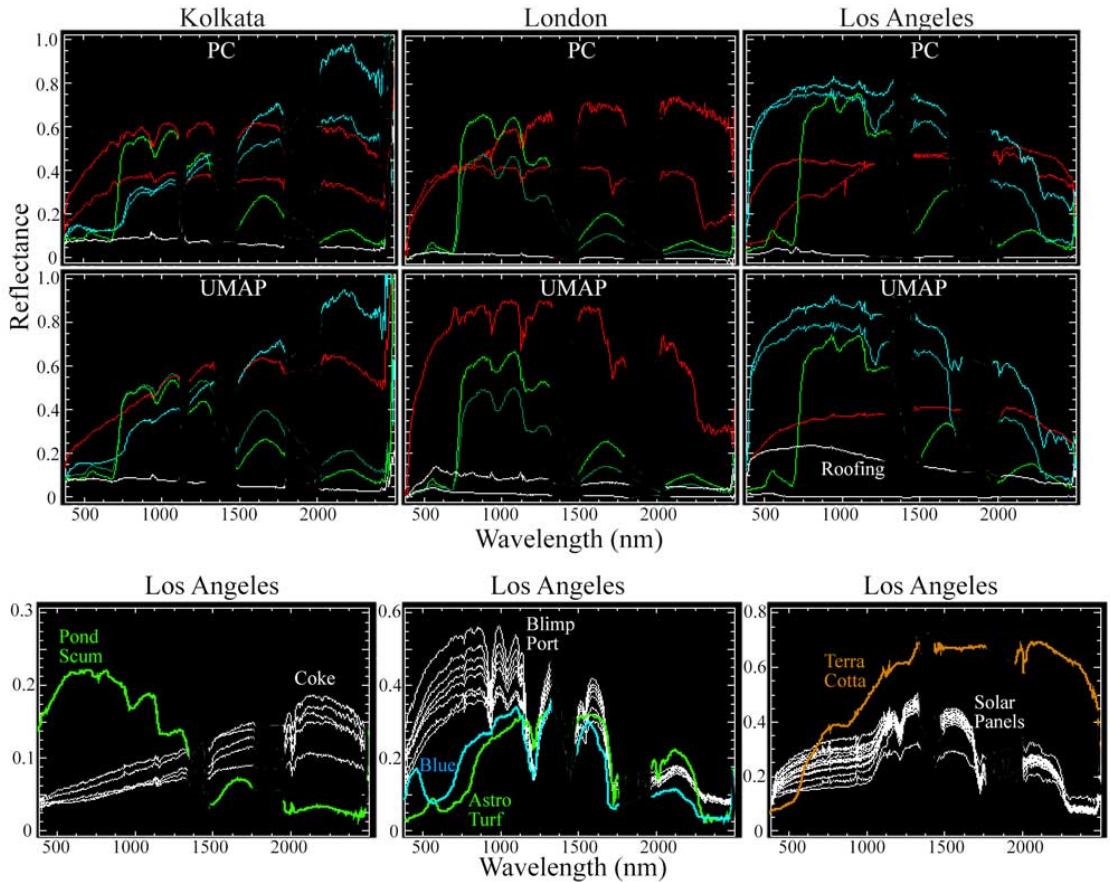

*Figure 9 Endmembers and anomalous spectra from AVIRIS-NG mixing spaces. Interior and exterior PC space endmembers (top) and UMAP apex endmembers (center) corresponding to labels on Fig. 8. Red and cyan spectra correspond to Substrates and Paints respectively. Vegetation endmembers (green) are virtually identical in both spaces because vegetation rarely has exterior endmembers. In contrast, Substrates and Paints can vary widely based on composition, BRDF and condition. The much greater spectral diversity of the Los Angeles mixing space results in much greater numbers of small clusters and fine tendrils outside the SVD continuum of the UMAP space in Fig. 8. Example spectra from these anomalous clusters and tendrils (bottom) have distinct continuum shapes and absorption features that are not accommodated by the SVD mixing space. The spectrum labeled Blue corresponds to a synthetic surface on a soccer pitch.*

When the 3D UMAP embedding of the EMIT mosaic is back-projected into geographic space as an RGB composite the geographic specificity of the embedding is apparent (Fig. 10). While intra-scene land cover variations are apparent as slightly different hues, inter-scene geographic differences are immediately apparent as distinctly different hues for different cities. The cities with two subscenes each (Bangkok, Los Angeles, New Delhi, New York, Tokyo) each show distinct hues from each other but similar hue distributions among each pair. In each of these cities, intra-scene hue variability is greater than inter-scene variability, indicating that fine scale heterogeneity of land cover reflectance is supplemented by broad scale homogeneity of reflectance distributions.

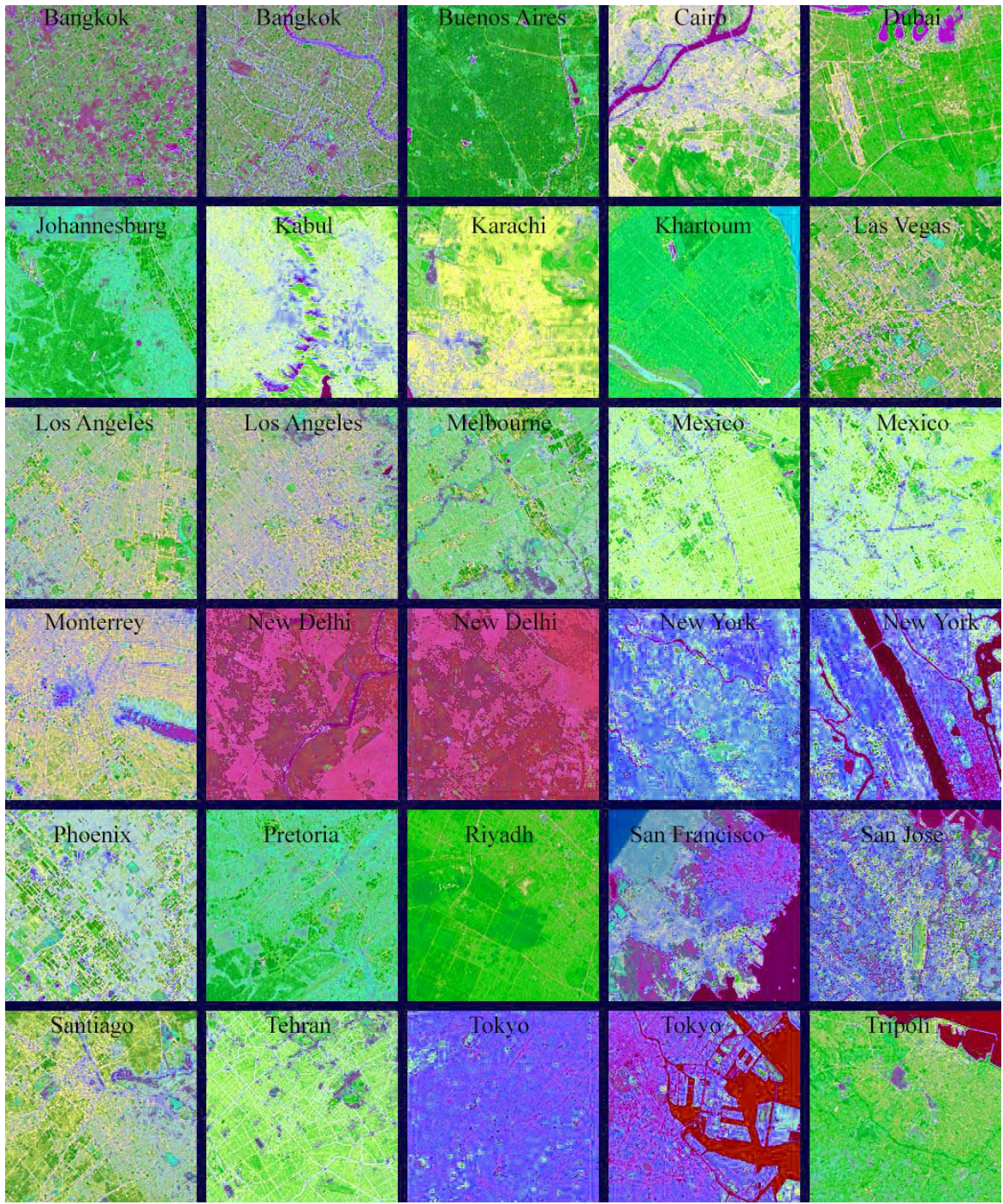

*Figure 10 UMAP embedding composites derived from a 3D embedding of the 30 subscene mosaic. Each RGB channel corresponds to a single UMAP dimension. EMIT spectra clustering together in UMAP space appear as similar colors. Despite intra-scene spectral diversity, same-scene pairs (e.g. Bangkok, Los Angeles, Mexico, New Delhi, New York, Tokyo) each share similar color palettes, suggesting shared intra-urban spectral properties, but distinct from city to city. Compare to the Vis/NIR/SWIR composite in Fig. 2. UMAP embeddings reflect the entire spectral mixing space topology.*

**Discussion**

*Spectral Dimensionality*
The consistency of the EMIT and AVIRIS-NG variance partition suggests that the spectral diversity of built environments approaches the spectral diversity of global compilations of spectrally heterogeneous landscapes – in both multispectral and spectroscopic feature spaces. This is consistent with the fact that variance-based dimensionality reduction like principal components are primarily sensitive to differences in the amplitude and curvature of the spectral continuum, which is generally resolved equally well by broadband and narrowband sensors. In contrast, non-linear manifold learning-based dimensionality reduction can preserve local topology related to low variance features like narrowband absorptions – while still preserving some global scale topology for comparison with variance-based mixing space topology. This allows the UMAP mixing spaces to resolve distinct clusters of anomalous spectra, like subtle differences in water body reflectance. This is particularly noteworthy in the Los Angeles AVIRIS-NG mixing space. Estimates of the intrinsic spectral dimensionality of embeddings like UMAP are beyond the scope of this study, but is the focus of active research efforts.

The use of RPCA to separate low rank and sparse components of the spectral mixing space shows promising results. The fact that the variance partition and topology of the low rank mixing space are so similar to that of the traditional L2 PC space bodes well for the quality of EMIT radiance measurements and the atmospheric corrections used to produce the Level 2 reflectance product. The difference in the spectral standard deviations of the low rank and sparse components (Fig. A1) clearly indicates that the sparse component is dominated by high amplitude spikes adjacent to the SWIR water absorption bands, and the presence of thermal emission spectra in at least one urban core. It is also worth noting that we intentionally did not apply the provided atmospheric correction masks provided with every EMIT scene. Nonetheless, the total variance of the sparse component is negligible (< 4%)compared to the variance of the low rank component. In two cases, large spatially coherent features were observed in the sparse component maps. These correspond to semi-translucent fog in the western part of the San Francisco subscene and nearly transparent atmospheric effects in the northwest part of the New Delhi subscene. The latter is responsible for the two distinct vegetation+impervious mixing continua observed for New Delhi in the UMAP mixing space (Figs. 5 & 6). Taken together, these observations suggest that RPCA may be useful as an anomaly identifier for EMIT and other spectroscopic image data.

*Implications of Substrate Mixing Space Topology*
In strong contrast to our parallel analysis of crystalline substrates *(Small and Sousa, In Prep.)*, the topology and dimensionality of the impervious substrate mixing spaces reflect the pervasive subpixel mixing that is endemic to built environments. Specifically, the diversity of substrates, presence of vegetation and ubiquitous shadow. Whereas the sands and evaporites that comprise the crystalline substrate mixing space are spectrally and compositionally distinct from each other, the compositional homogeneity that results from the common geologic provenance and sedimentologic refinement of sands and evaporites in each subscene produces distinct mixing continua in the PC space and very distinct clusters in the UMAP space. In contrast, the great diversity of impervious substrates used in built environments converges to a similar 3D SVD space to that found for spectrally diverse aggregations of both multispectral *(Small 2004; Small* *and Sousa 2022; Sousa and Small 2017a)* and hyperspectral *(Sousa and Small 2023, 2017b)* imagery. The notable difference is the presence of multiple Substrate endmembers, with both interior apexes and exterior constellations of spectra in the EMIT mixing space, as well as the sub-decameter AVIRIS-NG mixing spaces. As might be expected, the finer spatial resolution of the sub-decameter imagery resolves a greater variety of anomalous spectra outside the SVD mixing continuum – particularly for the more diverse land cover of the Los Angeles mosaic.

*Distinct Continua*
The combination of EMIT's high spectral resolution and UMAP's multiscale topology preservation reveals a number of distinct mixing continua corresponding to individual geographic locations. The implication is that compositional consistencies in different built environments can be resolved by the EMIT sensor, even in the presence of endemic compositional heterogeneity and pervasive spectral mixing within the 60 m IFoV. These consistencies are apparent in the comparisons of spectral continua in Fig. 6. Both spectral continuum shape and narrowband absorptions are clearly resolved between different geographic locations. The geographic comingling within the Dubai+Riyadh and Kabul+Tehran continua suggest either the use of similar types of building materials or the presence of spectrally similar dust mineralogy for each of these pairs of cities. Or perhaps both. The source(s) of this comingling are the focus of separate study. While the spectral separability of these impervious substrates bodes well for spectroscopic compositional mapping and thematic classification of built environments, the spectral separability of each built environment from its surrounding (and interspersed) pervious substrates remains an open question. Particularly in the presence of pervasive eolian dust deposition in arid and semiarid built environments.

*Complementary Projections of Spectroscopic Mixing Spaces*
The multiscale mixing space topology revealed by the combination of global scale spectral continuum topology and local scale absorption feature consistency provides a physically interpretable topology that can inform continuous spectral mixture models, as well as feature space clustering that can inform discrete thematic classifications. While this analysis did not include an explicit Joint Characterization *(Sousa and Small 2021)* of the EMIT or AVIRIS-NG mixing spaces, the similar topologies of the mixing spaces suggests that Joint Characterization of both spectral mixing spaces and mixture residual spaces may yield further constraints for both continuous and discrete models of pervious and impervious substrates in built environments.

*Implications for Mapping Built Environments*
Permeability is a hydraulic property, not a spectral property. To the extent that hydraulically impervious surfaces are a defining physical characteristic of built environments, mapping them on the basis of spectral reflectance alone seems to be a poorly posed problem. Comparative analyses of built environments using decameter-resolution multispectral imagery reveal pervasive subpixel mixing of a wide range of substrates, shadow and vegetation, suggesting that spectral heterogeneity may be the most consistent spectral property of built environments *(Small 2005, 200*9*)*. The narrowband spectroscopic results presented here support these earlier findings and extend their generality to narrowband spectra. Most of the urban cores used in this analysis contain varying amounts of vegetation, and in a few cases (e.g. Kabul, Khartoum, Monterrey, Mexico City, Riyadh) exposed pervious substrate in undeveloped areas. However, the selection of subscenes was chosen to maximize the amount and diversity of impervious substrates. Hence the focus on urban cores where pervious substrates are relatively rare.  The resulting mixing spaces, at both decameter and sub-decameter scales, reveal the presence of several distinct mixing continua with different spectral continuum shapes and narrowband SWIR2 absorptions when local topology is preserved (Fig. 5, 6).  This, and the presence of multiple substrate endmembers in the both the decameter EMIT and sub-decameter AVIRIS-NG mixing spaces, indicate that a single generic SVD mixture model is unlikely to be very effective for mapping impervious surfaces globally.  This may also explain why even a thoroughly trained convolutional neural network classification like the Dynamic World product has difficulty distinguishing built environments from pervious substrate classes with a wide range of class probabilities *(Small and Sousa 2023*).  Given the water retention properties of pervious substrates, and the time-varying effect of moisture darkening, temporal stability of reflectance may be a more diagnostic physical property for impervious surfaces in built environments than metrics computed from single date imagery *(Small 2019; Small et al. 2014)*.  In the context of spectroscopic mapping of built environments, this suggests two avenues for future research on the topic.  Comparisons of pervious and impervious substrate reflectance along built density gradients in different settings could allow for characterization of spectral contrast between spatially adjacent pervious and impervious substrates.  A topic that has received less attention is related to spectral masking.  The effect of dust coatings on impervious surfaces is relevant in many arid and semi-arid environments with intermittent rainfall.  Multitemporal spectroscopic analysis of impervious substrate reflectance could provide insight into the spectral effects of dust accumulation rate on different types of impervious substrate.  The ability of EMIT reflectance products to distinguish multiple spectral continuum shapes and narrowband absorption features in a diverse collection of impervious substrates and spectral mixing continua suggests that it, and similar imaging spectrometers, could be a useful tool for addressing both of these characteristics of built environments worldwide.

**Appendix A – Robust Principal Component Analysis**

Robust Principal Component Analysis (RPCA *(Candès et al. 2011)*) treats an image series *M* as the sum of a low-rank matrix *L* and a sparse matrix *S*:

$$M = L + S$$

The *L + S* matrix decomposition is generally ill-posed (NP-hard), but under weak assumptions *(Candès et al. 2011)* show that both *L* and *S* can be recovered exactly through a convex optimization referred to as Principal Component Pursuit. By optimizing the sum of the nuclear norm of *L* ($L_*$, sum of the singular values) and the weighted $L_1$ norm of *S*, the objective is to

$$minimize \quad \|L\|_* + \lambda \|S\|_1$$

$$subject\ to \quad L + S = M$$

where

$$\lambda = \frac{1}{\sqrt{n_{(1)}}}, n_{(1)} = \max(n_1, n_2)$$

and *L* is a general rectangular matrix of dimensions $n_1$ x $n_2$. One benefit of the RPCA optimization is that no tuning parameters are required. RPCA has been used for a variety of image analysis tasks, including image restoration, hyperspectral image denoising, facial recognition, multifocal imaging, spatiotemporal change detection, and video processing *(Bouwmans et al. 2018)*.

In this analysis, we apply RPCA to the 30 subscene mosaic of EMIT spectral cubes, then apply traditional L2 PCA to both the L and S component cubes to quantify variance partition (dimensionality) and mixing space topology for each. The variance partitions, given by the eigenvalues of the Singular Value Decomposition, show that the L component is effectively 3D, while the S component is at least 6D with greater residual variance dispersed over the higher dimensions (Fig. A1a). Computing the spectral standard deviation of L at each wavelength shows spectral variability similar in form to impervious substrate spectra varying in amplitude as a result of varying illuminations (Fig. A1b). In contrast, spectral standard deviation of the S component shows negligible variability at most wavelengths except adjacent to SWIR water absorptions. This suggests that the RPCA effectively separates the low rank structure of the spectral mixing space from narrowband noise resulting from the atmospheric correction in anomalous spectra. The topology of the S component PC space (Fig. A2) is dominated by distinct high amplitude limbs with endmembers corresponding to distinct features of the anomalous spectra.

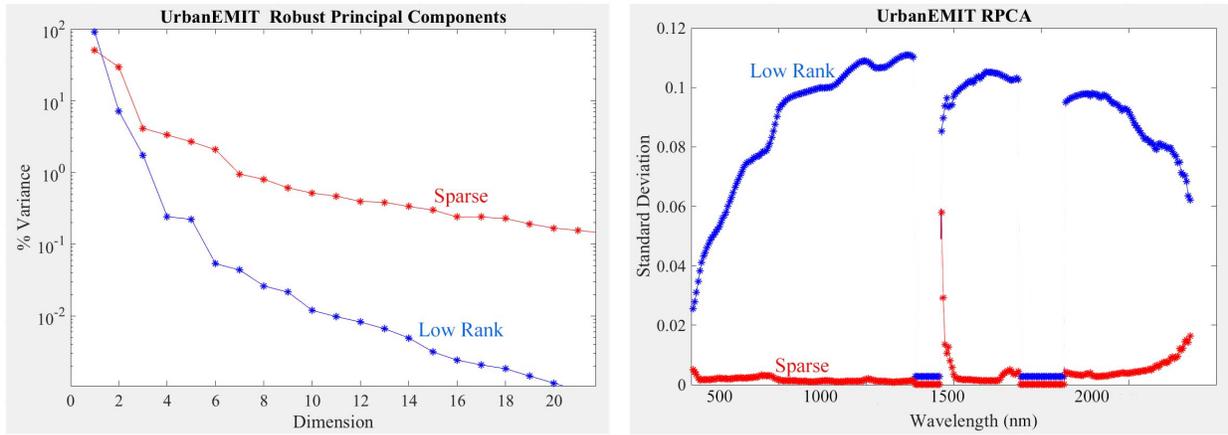

*Figure A1 Variance partition and spectral variability for Low Rank and Sparse components of the Urban EMIT mosaic. The Low Rank component feature space is effectively 3D, with > 99% of variance in 3 low order dimensions and < 1% in all higher. In contrast, the Sparse component is at least 6D with 92% of variance in 6 low order dimensions ad a much more gradual decay to higher dimensions. Standard deviation of all Low Rank spectra reflect amplitude variability of impervious substrate continua under varying illuminations. In contrast, spectral standard deviation of Sparse component spectra is negligible except adjacent to water absorption bands and in SWIR2.*

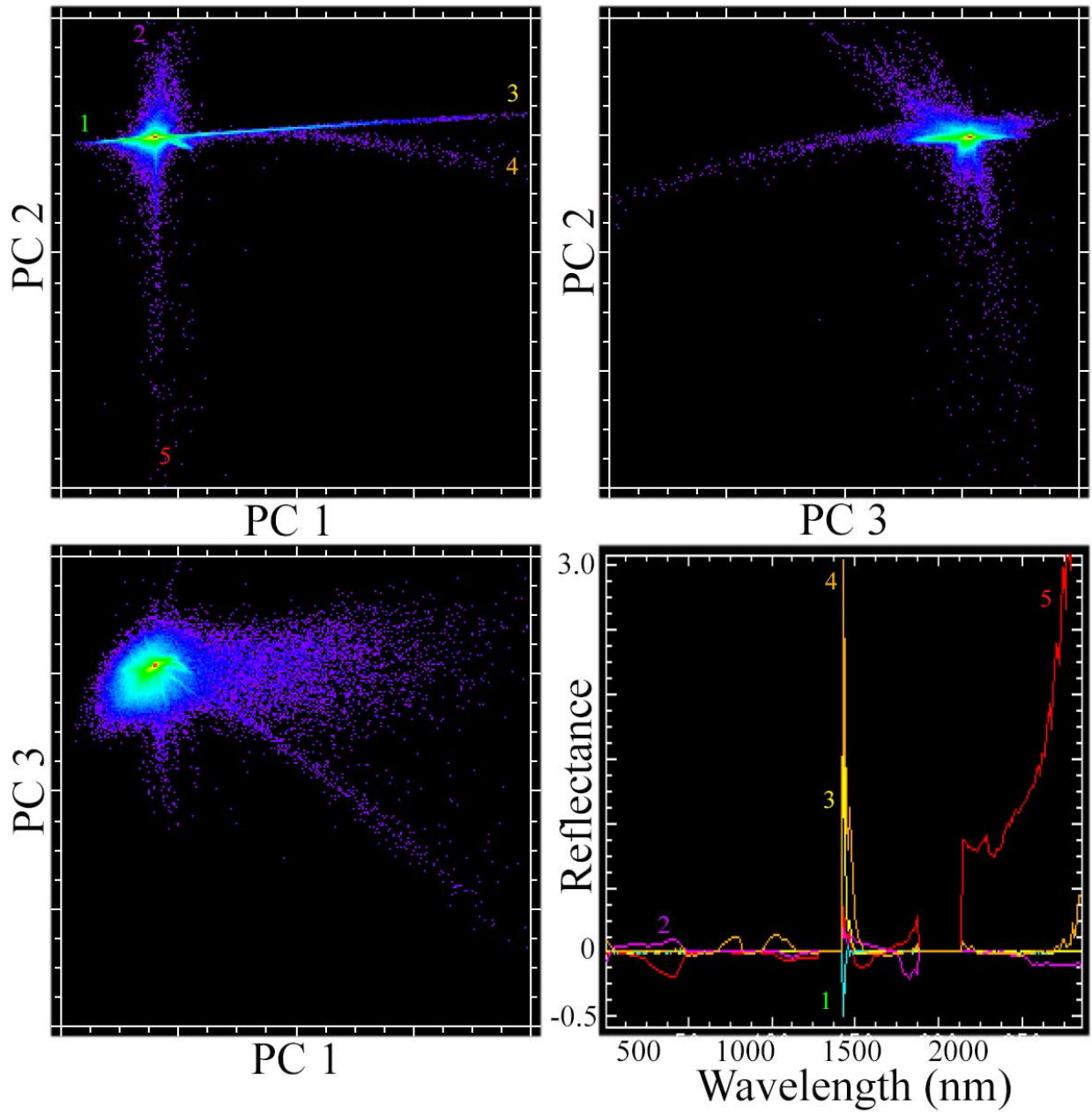

*Figure A2  3D spectral feature space and endmembers for the Sparse component of the Urban EMIT mosaic.  Low order principal component space reveals 5 prominent limbs corresponding to both high and low amplitude spectral features.  Aside from the thermal emission spectra (e.g. 5), most high amplitude features are at the edge of prominent absorption bands at SWIR wavelengths.*

**Appendix B – UMAP Sensitivity Analysis**

While the UMAP manifold learning algorithm has several hyperparameters, the topology of the resulting embeddings are generally most sensitive to the n_neighbors (NN) and min_dist (MD) settings (https://umap-learn.readthedocs.io/en/latest/).  NN controls how UMAP recovers the global vs local topology of the manifold, with higher values preserving more global structure at the expense of local detail. MD controls how UMAP projects the density of individual points in the low dimensional embedding, with lower values producing more distinct clusters. Hyperparameter sweeps of the EMIT mosaic show rapid convergence of mixing space topology for NN settings > 5, with similar numbers of clusters and tendril continua for all MD settings (Fig. A3). Hyperparameter sweeps for each of the AVIRIS-NG mosaics also show similarly rapid convergence for NN settings > 5, with negligible sensitivity to MD setting (Fig. A4) and similar SVD continua topology for each individual mosaic.

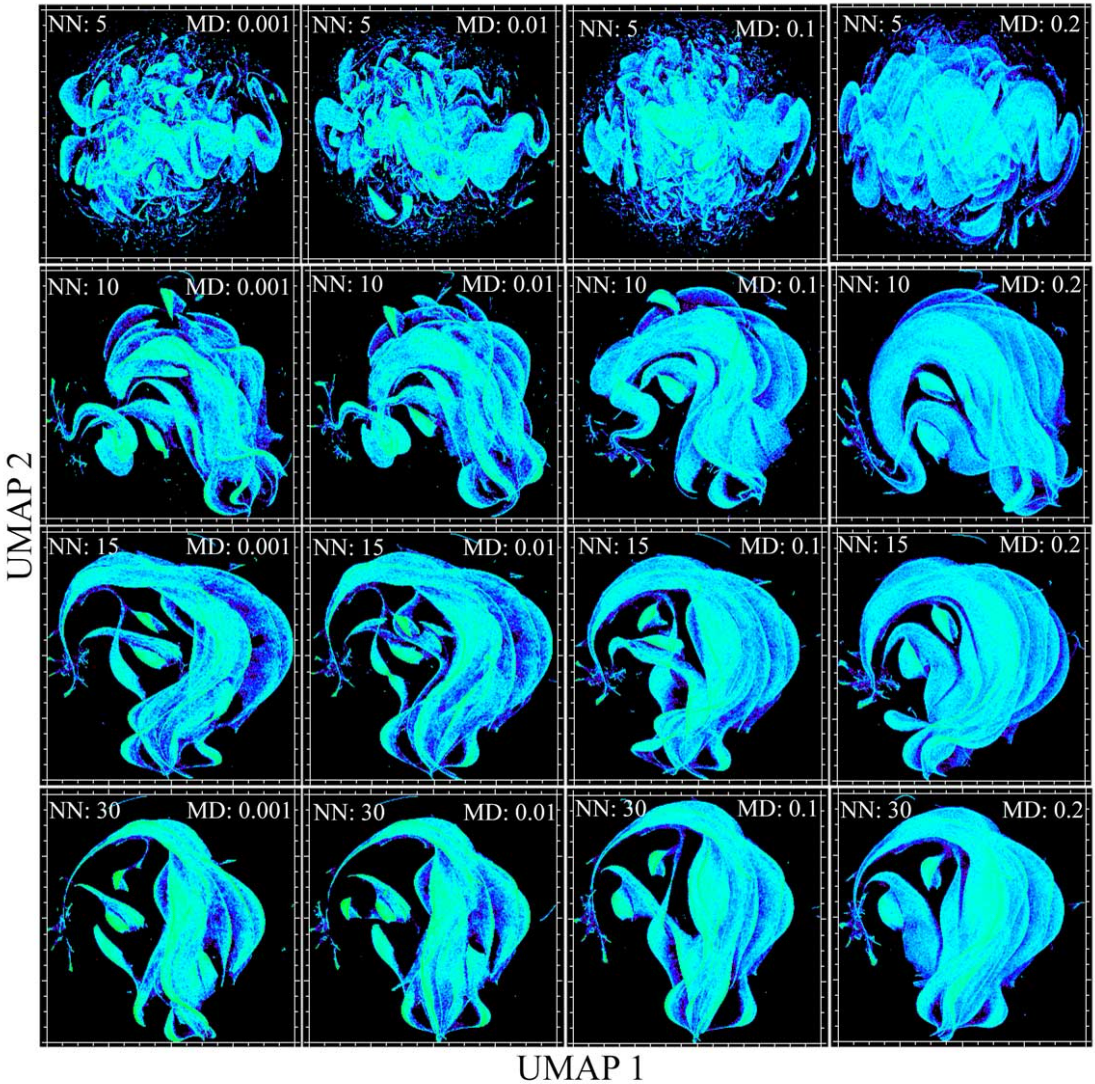

*Figure A3 Hyperparameter sweep for 2D UMAP embeddings of the urban EMIT mosaic. Varying Near_Neighbor (NN) and Minimum_Distance (MD) settings yield a consistent mixing space topology for all NN > 5. This topology takes the form of a set of parallel but distinct mixing continua spanning a few background Substrate reflectances and a larger number of impervious continua converging to a single Dark endmember surrounded by a constellation of water body clusters. Distinct from these Substrate - Impervious - Dark continua are multiple clusters corresponding to Impervious - Vegetation mixtures in cities containing unusually high abundances of urban vegetation.*

**Appendix C – Computational Specifics**

All computation was performed using commercial off-the-shelf Apple Macintosh hardware. Python and R runs were split between a 2020 MacBookPro (2 GHz Quad-Core Intel Core i5) and a 2019 iMac (3.1 GHz Hex-Core Intel i5). UMAP was run in a Python version 3.9.7 coding environment. Computation was executed using version 0.5.2 of the umap-learn package. The UMAP min_dist and n_neighbors hyperparameters were systematically varied as explained in the main text and Appendix B. The n_components hyperparameter was set to 2 or 3, depending on the desired dimension of the embedding space. The Euclidean distance metric was used for all runs. UMAP runtimes were generally in the range of 0.5 to 3 hours.

RPCA was run in an R version 4.2.0 coding environment. Computation was executed using verion 0.2.3 of the rpca package. RPCA hyperparameters were set as follows:
```
lambda = 1/sqrt(max(dim(M)))
mu = prod(dim(M))/(4 * sum(abs(M)))
term.delta = 10^(-7)
max.iter = 5000
trace = TRUE
thresh.nuclear.fun = thresh.nuclear
thresh.l1.fun = thresh.l1
F2norm.fun = F2norm
```

RPCA runtimes varied considerably, but were generally < 24 hours.

All L2 principal component transforms and associated Singular Value Decompositions were computed in the IDL+ENVI environment running on a 2015 iMac (3.3 GHz Quad-Core Intel i5), generally taking < 2 minutes.